%% file: paper_draft_v3.tex
\pdfoutput=1
\documentclass[conference,a4paper,10pt]{IEEEtran}

\usepackage[T1]{fontenc}
\usepackage{cite}
\usepackage[pdftex]{graphicx}
\graphicspath{{./figures/}}
\usepackage{empheq}
\usepackage{cases}
\usepackage[cmintegrals]{newtxmath}
\usepackage{bm}

\usepackage{algorithm}
\usepackage[noend]{algpseudocode}
\usepackage{dblfloatfix}
\usepackage{microtype}
\usepackage{url}
\usepackage{color}

\DeclareMathAlphabet{\pazocal}{OMS}{zplm}{m}{n}
\newcommand{\Ob}{\pazocal{O}}
\newcommand{\Pb}{\pazocal{P}}

\DeclareMathOperator*{\argmin}{\arg\!\min}

\algnewcommand{\IIf}[1]{\State\algorithmicif\ #1\ \algorithmicthen}
\algnewcommand{\EndIIf}{\unskip\ \algorithmicend\ \algorithmicif}

\usepackage[a4paper,top=1.905cm,bottom=3.2cm,left=1.42cm,right=1.42cm]{geometry}
\usepackage[pdfencoding=auto, psdextra]{hyperref}
\hypersetup{pdfauthor={Palaiologos, Michail; Castãneda~García, Mario H.; Laas, Tobias; Stirling-Gallacher, Richard A.; Caire, Giuseppe},pdftitle={Joint Antenna Selection and Covariance Matrix Optimization for ISAC Systems}}

\usepackage{etoolbox}
\usepackage{tikz}
\usetikzlibrary{tikzmark}
\usetikzlibrary{calc}

\errorcontextlines\maxdimen

\newcommand{\ALGtikzmarkcolor}{black}
\newcommand{\ALGtikzmarkextraindent}{4pt}
\newcommand{\ALGtikzmarkverticaloffsetstart}{-.5ex}
\newcommand{\ALGtikzmarkverticaloffsetend}{-.5ex}
\makeatletter
\newcounter{ALG@tikzmark@tempcnta}

\newcommand\ALG@tikzmark@start{%
	\global\let\ALG@tikzmark@last\ALG@tikzmark@starttext%
	\expandafter\edef\csname ALG@tikzmark@\theALG@nested\endcsname{\theALG@tikzmark@tempcnta}%
	\tikzmark{ALG@tikzmark@start@\csname ALG@tikzmark@\theALG@nested\endcsname}%
	\addtocounter{ALG@tikzmark@tempcnta}{1}%
}

\def\ALG@tikzmark@starttext{start}
\newcommand\ALG@tikzmark@end{%
	\ifx\ALG@tikzmark@last\ALG@tikzmark@starttext
	\else
	\tikzmark{ALG@tikzmark@end@\csname ALG@tikzmark@\theALG@nested\endcsname}%
	\tikz[overlay,remember picture] \draw[\ALGtikzmarkcolor] let \p{S}=($(pic cs:ALG@tikzmark@start@\csname ALG@tikzmark@\theALG@nested\endcsname)+(\ALGtikzmarkextraindent,\ALGtikzmarkverticaloffsetstart)$), \p{E}=($(pic cs:ALG@tikzmark@end@\csname ALG@tikzmark@\theALG@nested\endcsname)+(\ALGtikzmarkextraindent,\ALGtikzmarkverticaloffsetend)$) in (\x{S},\y{S})--(\x{S},\y{E});%
	\fi
	\gdef\ALG@tikzmark@last{end}%
}

\apptocmd{\ALG@beginblock}{\ALG@tikzmark@start}{}{\errmessage{failed to patch}}
\pretocmd{\ALG@endblock}{\ALG@tikzmark@end}{}{\errmessage{failed to patch}}
\makeatother

\usepackage{xcolor}
\usepackage{doi}

\begin{document}
	
\title{Joint Antenna Selection and Covariance Matrix Optimization for ISAC Systems}

\author{\IEEEauthorblockN{Michail Palaiologos\textsuperscript{$\ast$\dag}, Mario H. Cast\~aneda Garc\'ia\textsuperscript{$\ast$}, Tobias Laas\textsuperscript{$\ast$}, Richard A. Stirling-Gallacher\textsuperscript{$\ast$}\\
and Giuseppe Caire\textsuperscript{\dag}}
\IEEEauthorblockA{\textsuperscript{*}Munich Research Center, Huawei Technologies Duesseldorf GmbH, 80992 Munich, Germany\\
\IEEEauthorblockA{\textsuperscript{\dag}Communications and Information Theory Group, Technical University of Berlin, 10587 Berlin, Germany\\
Emails: \{\href{mailto:michail.palaiologos@huawei.com}{michail.palaiologos}, 
\href{mailto:mario.castaneda@huawei.com}{mario.castaneda},
\href{mailto:richard.sg@huawei.com}{richard.sg}\}@huawei.com;
\href{mailto:tobias.laas@tum.de}{tobias.laas@tum.de};
\href{mailto:caire@tu-berlin.de}{caire@tu-berlin.de}}}}
\maketitle%
\begin{tikzpicture}[remember picture,overlay]
\node[yshift=2cm] at (current page.south){\parbox{\textwidth}{\footnotesize \textcopyright\ 2024 IEEE. Personal use of this material is permitted. Permission from IEEE must be obtained for all other uses, in any current or future media, including reprinting/republishing this material for advertising or promotional purposes, creating new collective works, for resale or redistribution to servers or lists, or reuse of any copyrighted component of this work in other works.}};
\node[yshift=-1cm] at (current page.north){\parbox{\textwidth}{\footnotesize This is the accepted version of the following article: 
M.~Palaiologos, M.~H. Castãneda~García, T.~Laas, R.~A. Stirling-Gallacher, and G.~Caire, ``Joint antenna selection and covariance matrix optimization for {ISAC} systems,'' in \emph{Proc. IEEE Int. Conf. Commun. Workshops (ICC Workshops)}, Denver, CO, USA, Jun. 2024, pp. 2071--2076, \doi{10.1109/ICCWorkshops59551.2024.10615513}.}};
\end{tikzpicture}%
\begin{abstract}
We consider an integrated sensing and communication (ISAC) system with a single communication user and multiple targets. For the communication functionality, the achievable rate is employed as the performance metric, while for sensing, we focus on minimizing the mean squared error (MSE) between the designed beampattern and a desired one for tracking the targets. Towards this, and by assuming that there are fewer radio-frequency (RF) chains than antenna elements at the transmitter (Tx), we focus on the joint antenna selection (AS) and covariance matrix (CM) optimization at the Tx. This is a mixed-integer optimization problem, yet we demonstrate that it can be efficiently solved, in polynomial time, by combining convex optimization tools with dynamic programming (DP). By introducing an adjustable trade-off parameter, we formulate a joint objective function that captures both the communication and sensing metric. In this way, different ISAC solutions can be obtained, considering the trade-off among the two functionalities. It is shown that selecting the active antennas with our proposed method is superior than assuming a uniform Tx array with fixed antenna positions. Notably, by individually considering the optimization of either the sensing or the communication system alone, our proposed algorithm outperforms the literature proposals, by incurring only a small increase in complexity.
\end{abstract}

\IEEEpeerreviewmaketitle

\section{Introduction}\label{SecI}
Integrated sensing and communication (ISAC) systems have attracted a lot of attention, where communication and sensing functionalities operate on the same platform. By sharing the same hardware, operating frequencies and, possibly, waveform, ISAC systems offer tremendous benefits in terms of space, power consumption and spectral efficiency \cite{8999605}.

Specifically, multiple-input multiple-output (MIMO) ISAC systems have attracted considerable attention, as they can reap the benefits of both MIMO communication and MIMO radar systems \cite{8999605}. Namely, by means of spatial multiplexing, MIMO communication systems can offer a tremendous increase in the achievable rate of single and multi-user systems \cite{1192168}. MIMO radar systems can offer flexible design of the transmit beampattern by focusing the power around the targets' locations. In this way, power is not wasted in directions that are of no interest and, thus, the estimation performance of the targets' location parameters is drastically improved \cite{4350230}. One way of achieving this is by minimizing the mean squared error (MSE) between the designed beampattern and a desired one, while also minimizing the cross-correlation between the sensing (i.e., probing) signals at different target locations \cite{4276989}. 

Accordingly, to improve the performance of a MIMO ISAC system, several works propose transmit beamforming (BF) to leverage the spatial degrees of freedom \cite{10070799, WANG2018223, 10097184, 10243136}. In addition, antenna selection (AS) at the transmitter (Tx) is also employed as another means of enhancing the performance \cite{10070799, WANG2018223, 10097184, 10243136}. AS can be applied when there are more antenna elements than radio-frequency (RF) chains at the Tx, which enables the Tx array configuration to be adaptively configured. This is accomplished with a simple analog switching network, so that increased energy efficiency, reduced cost and hardware complexity can be attained with AS schemes \cite{10070799, 10243136}. 

In particular, to satisfy the signal-to-interference-plus-noise-ratio (SINR) constraints at the communication users and also achieve a suitable radar beampattern, joint transmit BF and AS is considered in \cite{10070799} by utilizing the alternating direction method of multipliers. In \cite{WANG2018223}, joint BF and AS is investigated towards minimizing the interference between sensing and communication, in the spatial domain, with sequential convex programming techniques. In \cite{10097184}, joint analog BF and AS is employed, by leveraging reinforcement learning, to focus the transmit power around the location of the sensing targets and communication users. In \cite{10243136}, a hybrid BF scheme is proposed, where the analog beamformer consists of phase shifters and switches, to simultaneously maximize the achievable rate at the communication users and the radar mutual information.

Evidently, to assess the performance of a MIMO ISAC system, the \textit{achievable rate} (or the SINR in multi-user systems) and the \textit{radar transmit beampattern} design are commonly used metrics. Notably, we point out that maximizing the achievable rate for a single-user MIMO communication system \cite{1192168} as well as minimizing the MSE between the designed and the desired transmit beampattern for a MIMO radar system \cite{4276989} can be achieved by optimizing the covariance matrix (CM) of the transmit signal. Moreover, AS has been widely considered for MIMO ISAC \cite{10070799, WANG2018223, 10097184, 10243136}, as well as, individually, for MIMO communication \cite{6030124, 6777403} and MIMO radar systems \cite{9020110, BOSE2021107985, 9376776}. 

Inspired by these, we consider joint AS and CM optimization at the Tx side towards improving the performance of an ISAC system. To the best of our knowledge, this particular problem has not been considered before. Namely, we assume a Tx that is using the same signal to transmit communication symbols to a user equipment (UE) and to probe several targets. To focus the power at the location of the targets, we aim at minimizing the MSE between the designed transmit beampattern and a desired one, while simultaneously maximizing the achievable communication rate at the UE. Towards this, we introduce a trade-off parameter in order to formulate a joint sensing and communication objective function.

Although the joint AS and CM problem is a mixed-integer optimization problem, hence difficult to solve, we demonstrate that by incorporating convex optimization tools \cite{boyd2004convex} with dynamic programming (DP) \cite{papadimitriou1998combinatorial}, it can be efficiently solved in polynomial time. Specifically, we demonstrate that:
\begin{itemize}
	\item A flexible performance trade-off between sensing and communication can be achieved with our proposed algorithm, so that the ISAC performance can be adjusted according to the system requirements.	
	\item The non-uniform Tx array, whose active antenna positions are selected with our algorithm, outperforms a fixed Tx array (i.e., without AS) in terms of both the beampattern MSE \textit{and} the achievable rate, indicating that AS can enhance the performance of the ISAC system.
	\item When the proposed problem is solved with respect to (wrt) one functionality only, our algorithm outperforms state-of-the-art proposals, both from the communication (wrt the achievable rate) \cite{6777403} and the radar literature (wrt the beampattern MSE) \cite{9376776}.
\end{itemize}

\textbf{Notation}: Bold lower case, upper case and non-bold letters represent vectors, matrices and scalars, respectively. $(\cdot)^{T}$ and $(\cdot)^{H}$ denote the transpose and conjugate transpose of a vector or matrix, respectively. The $i$-th element of $\mathbf{a}$ is denoted as $\text{a}_{i}$, while the $i$-th row and $j$-th column element of $\mathbf{A}$ as $\text{A}_{i,j}$. $E(\cdot)$ represents expected value and $\text{tr}(\mathbf{A})$ the trace of matrix $\mathbf{A}$. $\odot$ is the element-wise product and $\text{diag}(\mathbf{a})$ is a diagonal matrix formed from the elements of $\mathbf{a}$.

\section{System Model}\label{SecII}
We consider a MIMO ISAC system where a Tx array is using the \textit{same} signal to transmit communication symbols to a UE and to sense, i.e., probe $Q$ (already detected) targets. The reflected signals from the targets are captured by a sensing receive array, which can be collocated with the Tx (monostatic setup) or lie at a different physical location (bistatic setup) \cite{8999605}. The Tx and UE employ uniform linear arrays (ULAs) consisting of $N$ and $M$ antennas, respectively. There are $K$ RF chains at the Tx, where $K < N$. 

The transmitted signal during the $l$-th time slot is denoted as $\mathbf{x}(l) \in \mathbb{C}^{N}$, where $l = 1, \dots, L$ and $L$ is the number of samples/slots that each radar pulse/communication frame consists of. The CM $\mathbf{R} \in \mathbb{C}^{N \times N}$ of the transmit signal $\mathbf{x}(l)$ is given as \cite{1192168, 4350230}
\begin{equation}\label{eq:4}
	\mathbf{R} = E\left(\mathbf{x}(l) \mathbf{x}(l)^{H}\right).
\end{equation}
We assume that the total transmit power is equal to $P_{\text{Tx}}$, i.e., $\text{tr}(\mathbf{R}) \leq P_{\text{Tx}}$. As there are fewer RF chains than antenna elements at the Tx, only a subset of the available antennas can be selected each time. In light of this, we introduce a binary selection vector $\mathbf{p} \in \mathbb{B}^{N}$, whose entries correspond to particular antenna positions and are equal to 1 or 0, depending on the corresponding antenna being activated or not. As there are $K$ RF chains, the number of non-zero elements of $\mathbf{p}$ must be equal to $K$, i.e., $\sum_{n=1}^{N} p_{n} = K$.

\subsection{MIMO Communication Model}\label{SecIIa}
Regarding the communication model, the received signal $\mathbf{y}_{\text{c}}(l) \in \mathbb{C}^{M}$ at the UE at the $l$-th time slot reads as \cite{6030124}
\begin{equation}\label{eq:1}
	\mathbf{y}_{\text{c}}(l) = \mathbf{H}_{\text{c}} \mathbf{x}(l)  + \mathbf{n}_{\text{c}}(l), \quad l = 1, \dots, L.
\end{equation} 
$\mathbf{H}_{\text{c}} \in \mathbb{C}^{M \times N}$ is the channel matrix and $\mathbf{n}_{\text{c}}(l) \in \mathbb{C}^{M}$ is the complex additive white Gaussian noise at the UE, whose covariance matrix is $\mathbf{C} = \sigma_{\text{c}}^2 \mathbf{I}_{M}$. $\sigma_{\text{c}}^2$ is the noise variance and $\mathbf{I}_{M}$ is the $M \times M$ identity matrix. The communication channel is characterized by flat Rayleigh fading, i.e., the entries of $\mathbf{H}_{\text{c}}$ are circularly symmetric complex Gaussian variables, although other types of channel models can be also employed.

To optimize the communication performance, we consider maximizing the \textit{achievable rate} at the UE, which is given as a function of the transmit CM as \cite{1192168} 
\begin{equation}\label{eq:5}
C(\mathbf{R})  = \text{log}_2\left(\det \left(\mathbf{I}_{M} + \mathbf{H}_{\text{c}} \mathbf{R} \mathbf{H}_{\text{c}}^{H} \mathbf{C}^{-1}\right)\right).
\end{equation}
To incorporate the effect of the Tx AS, we formulate the diagonal matrix $\mathbf{\Delta} = \text{diag}(\mathbf{p}) \in \mathbb{B}^{N \times N}$, which acts as a binary selection matrix for the Tx antennas. Its $n$-th diagonal element $\Delta_{n, n}$ is equal to 1 or 0, depending on the Tx antenna located at position $n$ being activated or not, with $n = 1, \dots, N$. In this way, the achievable rate expression can be written as a function of both the CM and the AS matrix as \cite{6030124, 6777403}
\begin{equation}\label{eq:6}
C(\mathbf{\Delta}, \mathbf{R}) = \text{log}_2\left(\det \left(\mathbf{I}_{M} + \mathbf{H}_{\text{c}} \mathbf{\Delta} \mathbf{R} \mathbf{\Delta}^{H} \mathbf{H}_{\text{c}}^{H} \mathbf{C}^{-1}\right)\right).
\end{equation} 

\subsection{MIMO Radar Model}\label{SecIIb}
Wrt sensing, the goal of the Tx is to steer the probing signals towards the direction of the targets, as this would assist in the estimation of their location parameters \cite{4350230}. Towards this, we consider minimizing the MSE between the designed beampattern and a desired/optimum one, while also minimizing the cross-correlation of the probing signals at different target locations, as this would assist in the estimation process \cite{4276989}.

First, note that the power of the transmit signal $\mathbf{x}(l)$ at the direction of azimuth angle $\theta$ is equal to \cite{4276989}
\begin{equation}\label{eq:7}
P(\theta) = \mathbf{a}(\theta)^{H} \mathbf{R} \mathbf{a}(\theta),
\end{equation} 
where $\mathbf{R}$ is given by \eqref{eq:4} and $\mathbf{a}(\theta) \in \mathbb{C}^{N}$ is the Tx steering vector, given as
\begin{equation}\label{eq:77}
\mathbf{a}(\theta) = [1, e^{-j 2 \pi \frac{d_{\text{t}}}{\lambda} \sin \theta}, \dots, e^{-j 2 \pi \frac{(N - 1) d_{\text{t}}}{\lambda} \sin \theta}]^{T}.
\end{equation}
$\lambda$ is the wavelength and $d_{\text{t}}$ is the antenna spacing at the Tx ULA. Note from \eqref{eq:7} that the transmit beampattern is a function of the transmit signal's CM and the Tx array configuration. To take into account that only $K$ antennas can be selected, we can incorporate the binary selection vector $\mathbf{p}$ in \eqref{eq:7} as \cite{9020110, BOSE2021107985, 9376776}
\begin{equation}\label{eq:8}
P(\theta) = (\mathbf{p} \odot \mathbf{a}(\theta))^{H}  \mathbf{R} (\mathbf{p} \odot \mathbf{a}(\theta)).
\end{equation} 

Let $Q$ targets be located at angles $\{\theta_{q}\}_{q = 1}^{Q}$. Then, if the total number of angular scanning directions, i.e., of considered points, is equal to $G$, we can define a \textit{desired} beampattern over $\{\theta_{g}\}_{g = 1}^{G}$ as $P_{d}(\theta)$, whose shape should satisfy the requirement of maximizing the power around the locations $\{\theta_{q}\}_{q = 1}^{Q}$ of the targets \cite{4276989}. So, our goal is to design a beampattern $P(\theta)$ over all scanning directions $\{\theta_{g}\}_{g = 1}^{G}$ such that the MSE between $P(\theta)$ and $P_{d}(\theta)$ and also the cross-correlation at the two different locations $\theta_{q}$ and $\theta_{p}$, defined as $(\mathbf{p} \odot \mathbf{a}(\theta_{q}))^{H}  \mathbf{R} (\mathbf{p} \odot \mathbf{a}(\theta_{p}))$, are minimized. This can be expressed as \cite{9020110, BOSE2021107985, 9376776}
\begin{equation}\label{eq:9}
\begin{aligned}[b] & F \left(\mathbf{p}, \alpha, \mathbf{R}\right) = \frac{1}{G}{\sum \limits _{g = 1}^G \gamma_{g} \left|P(\theta_{g}) - \alpha P_{d}(\theta_{g})\right|^2} \\ &+ \frac{2 \omega}{Q^2 - Q} \sum \limits_{q = 1}^{Q - 1} {\sum \limits_{p = q + 1}^{Q} \left|(\mathbf{p} \odot \mathbf{a}(\theta_{q}))^H \mathbf{R} (\mathbf{p} \odot \mathbf{a}(\theta_{p}))\right|^2},
\end{aligned} 
\end{equation}
where $P({\theta_{g}})$ is given by \eqref{eq:8}. $\gamma_{g}$ denotes the weight (i.e., the importance) of each angular direction, $\omega$ is the weight of the cross-correlation term, while $\alpha$ is a scaling parameter, which must be also optimized according to the scaling of $P_{d}(\theta)$.  

\subsection{Problem Formulation}\label{SecIIc}
The goal of the ISAC system should be to maximize the achievable rate of the communication link between the Tx and the UE, while \textit{simultaneously} guaranteeing that the transmit power is properly focused towards the locations of the targets. To this extent, a joint metric that captures the ISAC system's performance can be formulated as
\begin{subequations}\label{eq:10}
	\begin{align} 
		\Pb_{1}: \ \underset{\mathbf{p}, \alpha, \mathbf{R}} {\text{min.}} \; &\ F \left({{\mathbf{p}}, \alpha, {\mathbf{R}}} \right) - \mu C(\mathbf{\Delta}(\mathbf{p}), \mathbf{R}),  \label{eq:10a} \\
		\text{s.t.:} &\ \; \mathbf{R} \succeq 0, \label{eq:10b} \\
		&\ \; \text{tr}(\mathbf{R}) \leq P_{\text{Tx}}, \label{eq:10c} \\
		&\ \; \mathbf{p} \in \mathbb{B}^{N}, \label{eq:10d} \\
		&\ \; \sum_{n = 1}^{N} \text{p}_{n} = K. \label{eq:10e}
	\end{align}
\end{subequations}
\eqref{eq:10b} and \eqref{eq:10e} respectively indicate that the CM must be positive semi-definite and that the number of simultaneously active Tx antennas must be equal to $K$. $\mathbf{\Delta}(\mathbf{p})$ in the achievable rate expression highlights the dependence of $\mathbf{\Delta}$ on the binary AS vector $\mathbf{p}$. Note the parameter $\mu \geq 0$ that is introduced in the objective function. By varying its value accordingly, a flexible \textit{performance trade-off} between the sensing and communication tasks can be enabled for the ISAC system.

Note also that a total power constraint is adopted in \eqref{eq:10c}, although a uniform power allocation is more practical for MIMO radar \cite{4276989}. However, as we consider an ISAC system, the power allocation is crucial for the communication performance, especially if the receive SNR at the UE is small \cite{6030124}. 

As $\Pb_{1}$ is a mixed-integer optimization problem, solving it is a very challenging task \cite{papadimitriou1998combinatorial}. Nevertheless, for a fixed AS vector $\mathbf{p}$, it is well-known that the expression of the achievable rate $C(\mathbf{\Delta}(\mathbf{p}), \mathbf{R})$ is \textit{concave} in $\textbf{R}$ \cite{1192168}. In addition, it was shown in \cite{4276989} that the expression of the beampattern MSE $F \left(\mathbf{p}, \alpha, \mathbf{R}\right)$ is (jointly) \textit{convex} in $\alpha$ and $\mathbf{R}$, provided that the binary AS vector $\mathbf{p}$ is fixed. Thus, for a fixed $\mathbf{p}$, the (weighted) difference $F \left({{\mathbf{p}}, \alpha, {\mathbf{R}}} \right) - \mu C(\mathbf{\Delta}(\mathbf{p}), \mathbf{R})$ between a convex and a concave function in the objective function of $\Pb_{1}$ is convex \cite{boyd2004convex}. Since the constraints \eqref{eq:10b} and \eqref{eq:10c} are also convex, we conclude that $\Pb_{1}$ is (jointly) \textit{convex} in $\mathbf{R}$ and $\alpha$, provided that $\mathbf{p}$ is fixed.

So, a straightforward approach for solving $\Pb_{1}$ would be an \textit{exhaustive search}. If each one of the $K$ RF chains at the Tx can connect to any of the $N$ available antennas, then for each one of the $\binom{N}{K}$ possible values of $\mathbf{p}$, $\Pb_{1}$ is solved over $\mathbf{R}$ and $\alpha$ (by, e.g., using the CVX optimization tool \cite{cvx}) and the global optimum is found. This has a prohibitively large complexity though, even for medium values of $N$ and $K$. Thus, in the following section we demonstrate how DP and convex optimization can be combined into an efficient algorithm towards obtaining a near-optimum solution to $\Pb_{1}$.

\section{Proposed Algorithm}\label{SecIII}
DP is a general purpose method that has been successfully applied for designing non-uniform arrays with low peak sidelobe level (PSL) \cite{1138163} and in combinatorial optimization \cite{papadimitriou1998combinatorial}. Yet, it has received very little attention in the context of AS for MIMO communications or radar\footnote{In \cite{BOSE2021107985} DP is also claimed to be used, however the proposed algorithm is based on a genetic algorithm and is entirely different than the algorithm we propose.}. We will show that it can lead to remarkable performance when used to solve $\Pb_{1}$. 

An AS algorithm must select the $K$ (out of the $N$ available) antennas that the $K$ RF chains must connect to. At each step, DP associates each candidate antenna that the $k$-th (for $k = 2, \dots, K$) RF chain is connected to with the optimal antenna that the previous, i.e., the $(k-1)$-th RF chain is connected to, by leveraging the fact that the associations between the previous $(k-1), \dots, 1$ RF chains have been already computed in previous steps \cite{1138163}. To explain how it works, we introduce the matrix $\mathbf{\Pi} \in \mathbb{N}^{N \times K}$ whose columns correspond to the $K$ RF chains and whose rows to the $N$ candidate antennas. The 1-st column is left blank and is only considered for simplifying the indexing. \textit{Each entry $\Pi_{n, k}$, for $n = 1, \dots, N$ and $k = 2, \dots, K$, contains the optimal antenna index (among the $N$ candidate ones) that the $(k-1)$-th RF chains is connected to, while the $k$-th RF chain is connected to the $n$-th antenna}. For example, $\Pi_{6, 4} = 5$ means that when the $k = 4$-th RF chain is connected to the $n=6$-th antenna, then the previous, i.e., the $(k-1)=3$-rd RF chain should be connected to the $\Pi_{6, 4} = 5$-th antenna.

So, starting with the 2-nd column of $\mathbf{\Pi}$, we sequentially consider all $N$ candidate antennas that the 2-nd RF chain can connect to by going over all $N$ row elements of this column. While this RF chain is connected to the, e.g., $n$-th antenna, we consider all $N$ candidate antennas that the \textit{previous} RF chain, i.e., the 1-st one, can connect to. Note that, while the 2-nd RF chain is connected to the $n$-th antenna, each connection of the 1-st RF chain to a different candidate antenna corresponds to a different configuration of $\mathbf{p}$, each one having only two non-zero elements. Given a configuration of $\mathbf{p}$, $\Pb_{1}$ can be easily solved with convex optimization tools. Thus, out of the $N$ antennas that the 1-st RF chain can connect to, the index of the antenna that leads to the best/minimum value of $\Pb_{1}$ is stored in the $(n, 2)$-th element of $\mathbf{\Pi}$, i.e., in $\Pi_{n, 2}$.

The remaining columns are filled in a similar way, considering that to fill the $k$-th column, we only need to evaluate all candidate antennas that the $(k-1)$-th RF chain can connect to, as the associations between previous RF chain pairs can be extracted with \textit{recursion} from the $(k-2), \dots, 2$ columns of $\mathbf{\Pi}$. Hence, an exhaustive search for jointly finding the best antenna indices that the $1$-st, $\dots$, $(k-1)$-th RF chain should be connected to is avoided, as the optimal antenna that previous RF chains must be connected to has already been determined and stored in the previous columns of $\mathbf{\Pi}$. So, by splitting the original problem into multiple sub-problems, DP allows to significantly reduce the computational complexity \cite{1138163}.

The process of filling $\mathbf{\Pi}$ is outlined in lines \ref{alg1:line1} - \ref{alg1:line13} of Alg. \ref{alg1}. In line \ref{alg1:line3}, $\mathbf{r}$ contains the indices of the antennas that the $1$-st, $\dots$, $k$-th RF chains are connected to. In lines \ref{alg1:line9} and \ref{alg1:line10} the already computed optimal antenna indices that the previous $(k - 2), \dots, 1$ RF chains should be connected to are obtained from the corresponding columns of $\mathbf{\Pi}$ by recursion. Since for a fixed $\mathbf{r}$, $\Pb_{1}$ is convex, its solution is stored in the elements of temporary vector $\mathbf{f}$ (line \ref{alg1:line11}). Also, $\{\text{p}(\mathbf{r}) = 1\}$ implies that the elements of $\mathbf{r}$ correspond to the elements of $\mathbf{p}$ that are set to 1 (for this iteration). Out of the $N$ candidate antennas that the $(k-1)$-th RF chain can connect to, the antenna index leading to the smallest solution of $\Pb_{1}$ is stored in $\Pi_{n, k}$ (line \ref{alg1:line12}).
\begin{algorithm}
	\caption{\textit{Solving $\Pb_{1}$ wrt $\mathbf{p} \in \mathbb{B}^{N}$, $\alpha$ and $\mathbf{R} \in \mathbb{C}^{N \times N}$}}\label{alg1}
	\begin{algorithmic}[1]
		\State $\mathbf{\Pi} = \mathbf{0}_{N \times K}$, $\mathbf{f}^{\prime} = \mathbf{0}_{N}$ \label{alg1:line1}
		\For {$k = 2 : K$}              \label{alg1:line2}
		\State $\mathbf{r} = \mathbf{0}_{k}$  \label{alg1:line3}
		\For {$n = 1 : N$}   \label{alg1:line4}
		\State $\text{r}_{k} = n$  \label{alg1:line5}
		\State $\mathbf{f} = \mathbf{0}_{N}$   \label{alg1:line6}
		\For{$n^{\prime} = 1 : N$}  \label{alg1:line7}
		\State $\text{r}_{k - 1} = n^{\prime}$ \label{alg1:line8}
		\For{$i = (k - 1) : -1 : 2$}	                \label{alg1:line9}
		\State $\text{r}_{i - 1}  = \Pi_{\text{r}_{i}, i}$     \label{alg1:line10}
		\EndFor
		\State $\text{f}_{n^{\prime}} = \Pb_{1}\{\text{p}(\mathbf{r}) = 1\}$   \Comment\textit{{Solve $\Pb_{1}$ wrt $\mathbf{R}$ and $\alpha$}}   \label{alg1:line11}
		\EndFor
		\State $\Pi_{n, k} = \mathop{\argmin} \, \mathbf{f}$     \label{alg1:line12}
		\IIf{$k = K$}  $\ \text{f}_{n}^{\prime} = \min \, \mathbf{f}$   \label{alg1:line13}
		\EndFor
		\EndFor
		\State $\text{r}_{K} = \text{argmin} \; \mathbf{f}^{\prime}$  \label{alg1:line14}
		\For {$i = K : -1 : 2$}  \label{alg1:line15}
		\State $\text{r}_{i - 1}  = \Pi_{\text{r}_{i}, i}$   \label{alg1:line16}
		\EndFor
		\State Solve $\Pb_{1}\{\text{p}(\mathbf{r}) = 1\}$   \Comment\textit{{Solve $\Pb_{1}$ wrt $\mathbf{R}$ and $\alpha$}}   \label{alg1:line17}
	\end{algorithmic}
\end{algorithm}

We emphasize that the \textit{final} $K$ selected antennas can be determined only after \textit{all} elements of $\mathbf{\Pi}$ have been computed. Towards this, for each antenna that the \textit{last}, i.e., the $K$-th RF chain is connected to, the smallest solution of $\Pb_{1}$ (out of the $N$ candidate ones) is stored in the elements of $\mathbf{f}^{\prime}$ in line \ref{alg1:line13}. The index of $\mathbf{f}^{\prime}$ with the minimum value corresponds to the optimal antenna that the $K$-th RF chain should connect to (line \ref{alg1:line14}). The optimal indices of the antennas that the $(K - 1)$-th, $\dots$, 1-st RF chains must connect to are obtained by recursion, i.e., by "moving backwards" through the columns of $\mathbf{\Pi}$ (lines \ref{alg1:line15} - \ref{alg1:line16}), since these contain all possible pairwise associations between the RF chains. The indices of all $K$ selected antennas\footnote{Although DP allows two RF chains to be connected to the same antenna, as this would result in inferior performance, it does not occur in practice.} are stored in the elements of $\mathbf{r}$.

After the $K$ selected antennas (i.e., the optimal configuration of $\mathbf{p}$) are obtained, $\Pb_{1}$ can be solved wrt $\mathbf{R}$ and $\alpha$ in line \ref{alg1:line17}, and the optimal CM is also obtained. We emphasize that the solution of Alg. \ref{alg1} is sub-optimal, since the AS problem cannot be split into sub-problems that are \textit{independent} of each other, that is, the principle of optimality of DP is not satisfied \cite{papadimitriou1998combinatorial}. Still, as we show below, this algorithm yields remarkable results when applied for solving $\Pb_{1}$.

Solving $\Pb_{1}$ wrt $\mathbf{R}$ and $\alpha$ with convex optimization tools with an accuracy of $\varepsilon$ amounts to a complexity of $\Ob \left(\log \left(\frac{1}{\varepsilon}\right) N^{3.5}\right)$ \cite{boyd2004convex}. As this needs to be solved $N^2 (K - 1) + 1$ times in Alg. \ref{alg1}, the total computational complexity of Alg. \ref{alg1} is $\Ob \left(\log \left(\frac{1}{\varepsilon}\right) N^{5.5} K\right)$, which is significantly lower than the complexity of exhaustive search, which is in the order of $\Ob (N!)$.

\section{Numerical Results}\label{SecIV}
The performance of the ISAC system is evaluated (with simulations carried out with MATLAB) in terms of the achievable rate (measured in bits per channel use (bpcu)) and of the beampattern MSE, when the joint AS and CM optimization problem $\Pb_{1}$ is solved with Alg. \ref{alg1}. For the Tx ULA spacing, the transmit power and the noise variance at the UE, we assume $d_{\text{t}} = \lambda/2$, $P_ {\text{Tx}} = 1$ and $\sigma_{\text{c}}^{2} = 0.01$. The angular scanning directions for sensing are within the range $[-90^{\circ}, 90^{\circ}]$ with a spacing of $1^{\circ}$, i.e., $G = 181$. All weights $\{\gamma_{g}\}_{g = 1}^{G}$ in \eqref{eq:9} are set to one, while $\omega = 1$. We assume $Q = 2$ targets, located at angles $-30^{\circ}$ and $30^{\circ}$. For a proper evaluation of our proposal, two sets of simulation results are considered. 

For the first one, we assume $N = 12$ antennas and $K=8$ RF chains at the Tx array, while the UE consists of $M=4$ antennas. The desired beampattern for sensing is given as
\begin{equation}\label{eq:12}
P_{d}(\theta) = \left\{ \begin{array}{rcl}
	1, & \theta \in  \left([-37^{\circ}, -23^{\circ}] \cup [23^{\circ}, 37^{\circ}]\right) \\ 0, & \text{otherwise}.
\end{array}\right.
\end{equation}
$P_{d}(\theta)$ consists of two mainlobes, which are centered at the locations of the targets, each one having a beamwidth of $15^{\circ}$. 

First, we consider the joint AS and CM optimization problem $\Pb_{1}$ with $\mu = [0, 0.0001, 0.001, 0.01, 0.1, 1]$. In this way, we can evaluate the communication and sensing operations, by considering the \textit{trade-off} between the achievable rate and the beampattern MSE when $\Pb_{1}$ is solved with our proposed Alg. \ref{alg1} (i.e., with DP) and with an exhaustive search (ES). We also consider the case where a \textit{fixed} ULA consisting of the same number of antennas and RF chains ($N = K = 8$) is employed at the Tx, i.e., all elements of $\mathbf{p}$ are fixed to 1 and $\Pb_{1}$ is solved wrt $\mathbf{R}$ and $\alpha$ with convex optimization tools. We highlight that by setting $\mu=0$ in $\Pb_{1}$, only the beampattern MSE is considered for the optimization, hence the system basically acts as a MIMO \textit{radar} system. In contrast, when $\mu=1$ in $\Pb_{1}$, only the communication functionality is considered (assuming $\mu > 1$ did not considerably affect the results), hence the ISAC system acts as a MIMO \textit{communication} system.

\begin{figure}[!t]
	\centering
	\def\svgwidth{0.9\columnwidth}
	\scalebox{1}{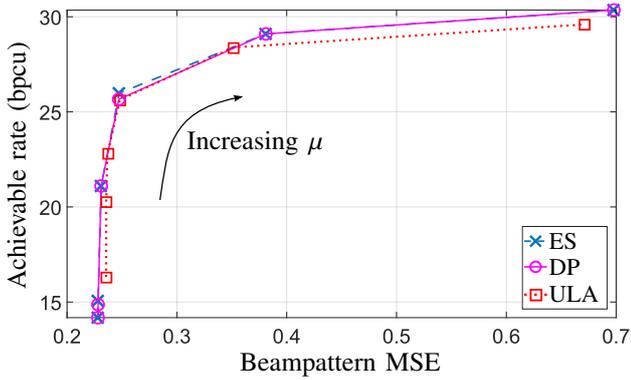}
	\caption{Achievable rate versus (vs.) beampattern MSE for $\mu = [0, 0.0001, 0.001, 0.01, 0.1, 1]$ for the \textit{first} setup.}
	\label{fig:tradeOff}
\end{figure}

The resulting trade-off curve is depicted in Fig. \ref{fig:tradeOff} (for illustration purposes, the points that correspond to different values of $\mu$ are connected). First, we notice that the performance of a system, where the selected antennas and the transmit CM are jointly optimized with our algorithm, is superior than the performance of a system where the Tx array configuration remains fixed and only the CM is optimized. In fact (although not directly visible from Fig. \ref{fig:tradeOff}) the optimal value of $\Pb_{1}$ in \eqref{eq:10a}, when it is solved with DP (wrt $\mathbf{p}$, $\mathbf{R}$ and $\alpha$), is smaller than the optimal value achieved with the fixed ULA, for all values of $\mu$. In addition, for most values of $\mu$, the performance of DP is \textit{identical} to the performance of the ES, despite the tremendous reduction in the computational complexity. So, Alg. \ref{alg1} can achieve the global optimum of $\Pb_{1}$ for some arrangements, thus resulting in an ISAC system that \textit{simultaneously} achieves low beampattern MSE and high communication rate. 

\begin{figure}[!t]
	\centering
	\def\svgwidth{0.9\columnwidth}
	\scalebox{1}{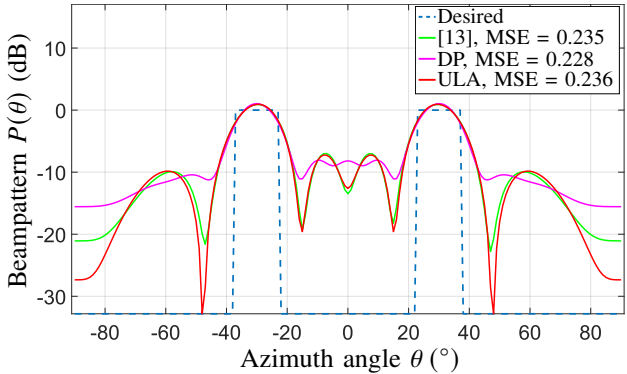}
	\caption{Beampattern vs. azimuth angle for the sensing-only optimization, i.e., for $\mu = 0$, for the \textit{first} setup.}
	\label{fig:Beampatt}
\end{figure}

Specifically, the performance can be easily adjusted according to the system requirements. For instance, when $\mu=0$, the rate in Fig. \ref{fig:tradeOff} is reduced by more than $50 \, \%$ compared to the achieved rate when $\mu=1$. For $\mu=0$, the resulting beampatterns are illustrated in Fig. \ref{fig:Beampatt}. These are given by \eqref{eq:8}, where $\mathbf{p}$ and $\mathbf{R}$ are optimized with Alg. \ref{alg1}, with ES, with a fixed ULA at the Tx (where $\mathbf{p}$ is fixed), as well as with the algorithm from \cite{9376776}. The latter is also focused on minimizing the beampattern MSE for a MIMO \textit{radar} system via joint AS and CM optimization. It offers the best performance and complexity (equal to $\Ob \left(\log \left(\frac{1}{\varepsilon}\right) N^{4.5}\right)$, given that, typically, $N >> K$) compared to related works \cite{9020110, BOSE2021107985}. Notably, the obtained beampatterns with ES and with Alg. \ref{alg1} are identical, so only the latter is depicted in Fig. \ref{fig:Beampatt}. Moreover, the resulting beampattern when Alg. \ref{alg1} is employed outperforms the beampattern from \cite{9376776}, as well as the one of the fixed ULA in terms of the beampattern MSE, by also achieving a lower PSL.

On the contrary, when $\mu=1$, the beampattern MSE is four times as high compared to $\mu=0$. We emphasize that a large MSE leads to poor estimation performance of the targets' location parameters \cite{4350230, 4276989}. For this first setup, the achievable rate is depicted in the \textit{second column} of Table \ref{table1}. This is given by \eqref{eq:6} and is obtained when $\Pb_{1}$ is solved (for $\mu=1$) with Alg. \ref{alg1}, with ES, with a fixed ULA at the Tx (where only $\mathbf{R}$ and $\alpha$ are optimized), and with the algorithm proposed in \cite{6777403}. 
\begin{table}[!t]
	\renewcommand{\arraystretch}{1.4}
	\caption{Achievable rate of different schemes for the communication-only optimization ($\mu=1$)}
	\label{table1}
	\centering
	\begin{tabular}{c||c||c}
		\hline
		\bfseries Scheme & \bfseries Rate (bpcu) -- 1st Setup & \bfseries Rate (bpcu) -- 2nd Setup  \\
		\hline\hline
		ES & 30.372 & 45.572 \\
		\hline
		\cite{6777403} & 28.830 & 43.308 \\
		\hline
		DP & 30.371 & 45.517 \\
		\hline
		ULA & 28.815 & 43.077 \\
		\hline\hline
	\end{tabular}
\end{table}

The algorithm from \cite{6777403} is also focused on maximizing the achievable rate by jointly optimizing the AS and CM for a MIMO \textit{communication} system. It is an iterative algorithm, whose complexity is $\Ob \left(\log \left(\frac{1}{\varepsilon}\right) N^{3.5} P\right)$. $P$ is the number of iterations, which depends on the required accuracy and is not known a-priori. For our results, we assumed that the algorithm converged when the difference in the rate between consecutive iterations was lower than 0.01 bpcu. So, from this second column of Table \ref{table1}, we notice that Alg. \ref{alg1} achieves almost the same rate as the ES, but a higher rate than the algorithm from \cite{6777403} and the fixed ULA with no AS at the Tx.

\begin{figure}[!t]
	\centering
	\def\svgwidth{0.9\columnwidth}
	\scalebox{1}{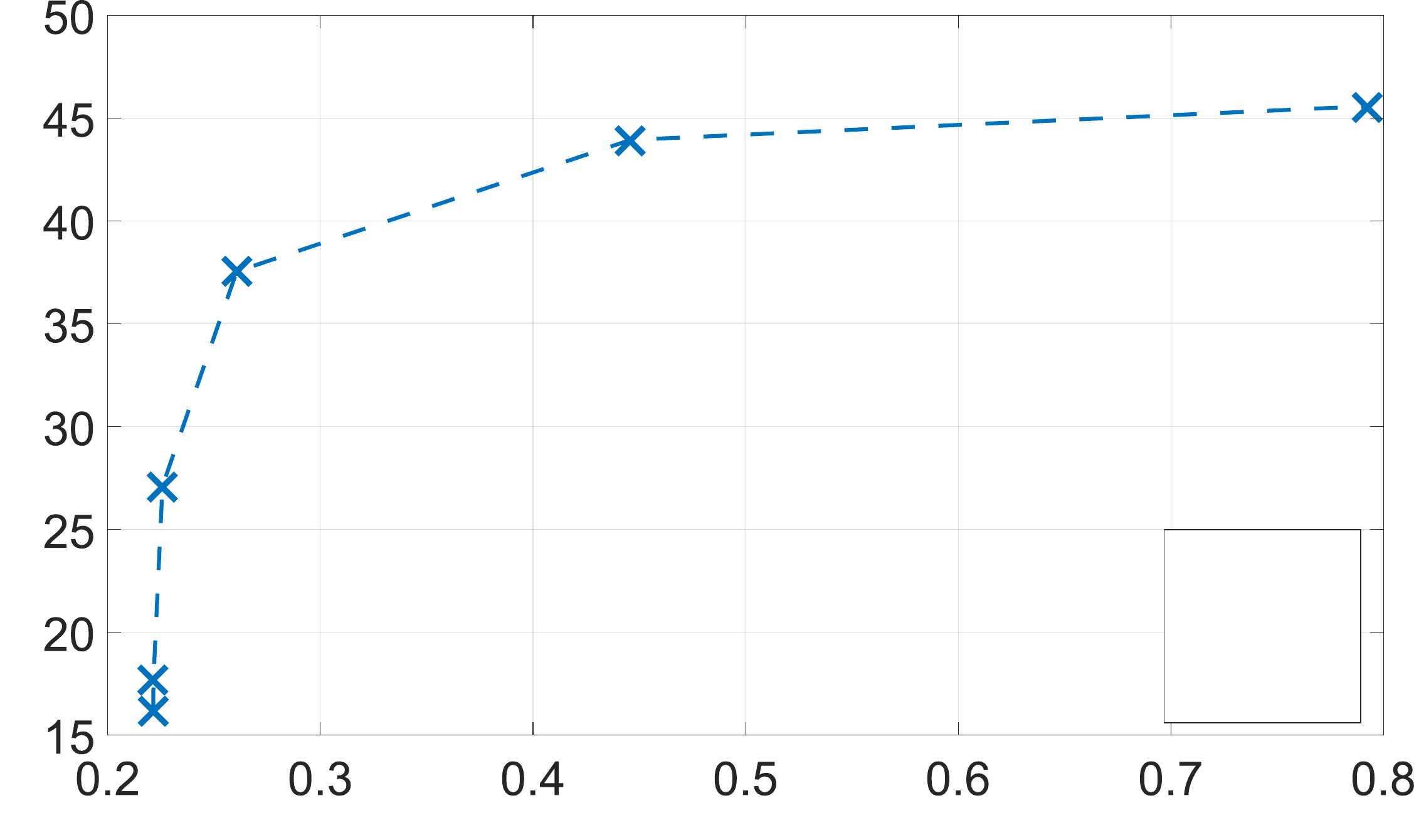}
	\caption{Achievable rate vs. beampattern MSE for $\mu = [0, 0.0001, 0.001, 0.01, 0.1, 1]$ for the \textit{second} setup.}
	\label{fig:BeampattULAs}
\end{figure}

To provide more insight, we also consider a second setup, where a larger number of antennas and RF chains is assumed, namely, $N = 20, \, K = 12$ and $M=6$. Also, the beamwidth of $P_{d}(\theta)$ around the targets' location is now equal to $11^{\circ}$ (i.e., narrower). Fig. \ref{fig:BeampattULAs} illustrates the performance trade-off of the system, when $\Pb_{1}$ is solved for varying values of $\mu$. We compare the performance of the proposed DP-based algorithm against the ES and a fixed ULA where there is no AS ($N=K=12$). 

Evidently, the achieved performance, when the joint AS and CM optimization problem is solved with our algorithm, is almost identical to the one of the ES, indicating that all Pareto optimal points can be approximately attained with our algorithm. Moreover, the performance of our proposal is superior than the performance with a fixed array at the Tx, that is, our algorithm achieves a better \textit{combination} of the rate and beampattern MSE values for all $\mu$. So, having more antennas than RF chains at the Tx and adaptively selecting the active ones can increase the performance of the ISAC system, by keeping the cost and hardware complexity low, since only a simple switching network is required to perform AS \cite{10070799, 10243136}.

\begin{figure}[!t]
	\centering
	\def\svgwidth{0.9\columnwidth}
	\scalebox{1}{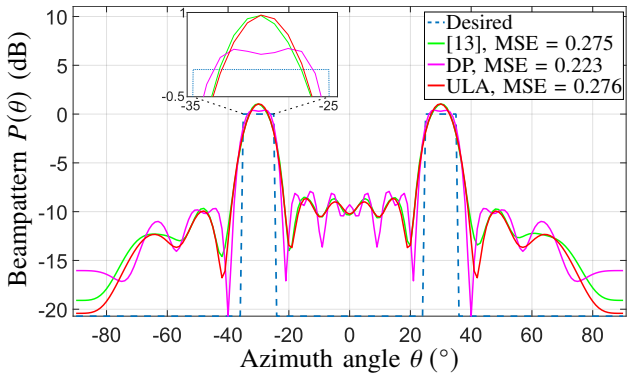}
	\caption{Beampattern vs. azimuth angle for the sensing-only optimization, i.e., for $\mu = 0$, for the \textit{second} setup.}
	\label{fig:Beampatt_bad}
\end{figure}

The ensuing beampatterns when only the sensing functionality is considered are depicted in Fig. \ref{fig:Beampatt_bad}. Our proposal better matches the desired beampattern compared to the proposal from \cite{9376776} and to the fixed Tx ULA. This is mainly evident in the region of the mainlobes, where the DP-based beampattern attains a flatter shape, indicating that the transmit power is better focused around the locations of the targets.

Finally, in the \textit{third column} of Table \ref{table1}, the achievable rate is shown when the system is optimized wrt the performance of the communication functionality only, i.e., when $\Pb_{1}$ is solved for $\mu=1$. The proposed algorithm outperforms the algorithm from \cite{6777403} and the fixed Tx ULA, in terms of the achievable rate, while the performance gap from the ES is minimal.

\section{Conclusions}\label{SecV}
We considered the joint antenna selection and covariance matrix optimization problem at the Tx towards optimizing an ISAC system. The single-user achievable rate and the MSE between the designed beampattern and a desired one were considered as the performance metrics, which were combined into a single objective function by introducing a trade-off parameter. To solve this difficult optimization problem, we proposed an algorithm which is based on combining dynamic programming with convex optimization tools. It was shown that a flexible performance trade-off between the communication and sensing functionality is attained with our algorithm, whose performance is comparable to that of the exhaustive search, despite the massive decrease in complexity. Moreover, our proposal outperforms a system with a fixed ULA at the Tx, indicating that antenna selection can provide performance enhancement, without the need for complicated analog and/or digital networks. Finally, considering the optimization wrt the communication or sensing functionality only, our algorithm outperforms state-of-the-art proposals, by admitting only a small increase in complexity.

\bibliographystyle{myIEEEtran}
\bibliography{IEEEabrv, library}

\end{document}

%% file: fig1_v4.pdf_tex
\begingroup%
  \makeatletter%
  \providecommand\color[2][]{%
    \errmessage{(Inkscape) Color is used for the text in Inkscape, but the package 'color.sty' is not loaded}%
    \renewcommand\color[2][]{}%
  }%
  \providecommand\transparent[1]{%
    \errmessage{(Inkscape) Transparency is used (non-zero) for the text in Inkscape, but the package 'transparent.sty' is not loaded}%
    \renewcommand\transparent[1]{}%
  }%
  \providecommand\rotatebox[2]{#2}%
  \newcommand*\fsize{\dimexpr\f@size pt\relax}%
  \newcommand*\lineheight[1]{\fontsize{\fsize}{#1\fsize}\selectfont}%
  \ifx\svgwidth\undefined%
    \setlength{\unitlength}{1078.83856201bp}%
    \ifx\svgscale\undefined%
      \relax%
    \else%
      \setlength{\unitlength}{\unitlength * \real{\svgscale}}%
    \fi%
  \else%
    \setlength{\unitlength}{\svgwidth}%
  \fi%
  \global\let\svgwidth\undefined%
  \global\let\svgscale\undefined%
  \makeatother%
  \begin{picture}(1,0.59967079)%
    \lineheight{1}%
    \setlength\tabcolsep{0pt}%
    \put(0,0){\includegraphics[width=\unitlength,page=1]{fig1_v4.pdf}}%
    \put(0.868,0.192){\color[rgb]{0,0,0}\makebox(0,0)[lt]{\lineheight{1.25}\smash{\small{\begin{tabular}[t]{l}ES\end{tabular}}}}}%
    \put(0,0){\includegraphics[width=\unitlength,page=2]{fig1_v4.pdf}}%
    \put(0.868,0.149){\color[rgb]{0,0,0}\makebox(0,0)[lt]{\lineheight{1.25}\smash{\small{\begin{tabular}[t]{l}DP\end{tabular}}}}}%
    \put(0.868,0.103){\color[rgb]{0,0,0}\makebox(0,0)[lt]{\lineheight{1.25}\smash{\small{\begin{tabular}[t]{l}ULA\end{tabular}}}}}%
    \put(0,0){\includegraphics[width=\unitlength,page=3]{fig1_v4.pdf}}%
    \put(0.358,-0.01){\color[rgb]{0,0,0}\makebox(0,0)[lt]{\lineheight{1.25}\smash{\begin{tabular}[t]{l}Beampattern MSE\end{tabular}}}}%
    \put(0,0){\includegraphics[width=\unitlength,page=4]{fig1_v4.pdf}}%
    \put(0.005,0.125){\color[rgb]{0,0,0}\rotatebox{90}{\makebox(0,0)[lt]{\lineheight{1.25}\smash{\begin{tabular}[t]{l}Achievable rate (bpcu)\end{tabular}}}}}%
    \put(0,0){\includegraphics[width=\unitlength,page=5]{fig1_v4.pdf}}%
    \put(0.27123877,0.35647019){\color[rgb]{0,0,0}\makebox(0,0)[lt]{\lineheight{1.25}\smash{\begin{tabular}[t]{l}Increasing $\mu$\end{tabular}}}}%
    \put(0,0){\includegraphics[width=\unitlength,page=6]{fig1_v4.pdf}}%
  \end{picture}%
\endgroup%

%% file: fig2_v2.pdf_tex
\begingroup%
  \makeatletter%
  \providecommand\color[2][]{%
    \errmessage{(Inkscape) Color is used for the text in Inkscape, but the package 'color.sty' is not loaded}%
    \renewcommand\color[2][]{}%
  }%
  \providecommand\transparent[1]{%
    \errmessage{(Inkscape) Transparency is used (non-zero) for the text in Inkscape, but the package 'transparent.sty' is not loaded}%
    \renewcommand\transparent[1]{}%
  }%
  \providecommand\rotatebox[2]{#2}%
  \newcommand*\fsize{\dimexpr\f@size pt\relax}%
  \newcommand*\lineheight[1]{\fontsize{\fsize}{#1\fsize}\selectfont}%
  \ifx\svgwidth\undefined%
    \setlength{\unitlength}{1072.28805542bp}%
    \ifx\svgscale\undefined%
      \relax%
    \else%
      \setlength{\unitlength}{\unitlength * \real{\svgscale}}%
    \fi%
  \else%
    \setlength{\unitlength}{\svgwidth}%
  \fi%
  \global\let\svgwidth\undefined%
  \global\let\svgscale\undefined%
  \makeatother%
  \begin{picture}(1,0.60192376)%
    \lineheight{1}%
    \setlength\tabcolsep{0pt}%
    \put(0,0){\includegraphics[width=\unitlength,page=1]{fig2_v2.pdf}}%
    \put(0.367,0.00){\color[rgb]{0,0,0}\makebox(0,0)[lt]{\lineheight{1.25}\smash{\begin{tabular}[t]{l}Azimuth angle $\theta \, (^{\circ})$\end{tabular}}}}%
    \put(0.014,0.13){\color[rgb]{0,0,0}\rotatebox{90}{\makebox(0,0)[lt]{\lineheight{1.25}\smash{\begin{tabular}[t]{l}Beampattern $P(\theta)$ (dB)\end{tabular}}}}}%
    \put(0,0){\includegraphics[width=\unitlength,page=2]{fig2_v2.pdf}}%
    \put(0.692,0.466){\color[rgb]{0,0,0}\makebox(0,0)[lt]{\lineheight{1.25}\smash{\footnotesize{\begin{tabular}[t]{l}ULA, MSE = 0.236\end{tabular}}}}}%
    \put(0,0){\includegraphics[width=\unitlength,page=3]{fig2_v2.pdf}}%
    \put(0.692,0.57){\color[rgb]{0,0,0}\makebox(0,0)[lt]{\lineheight{1.25}\smash{\footnotesize{\begin{tabular}[t]{l}Desired\end{tabular}}}}}%
    \put(0.692,0.536){\color[rgb]{0,0,0}\makebox(0,0)[lt]{\lineheight{1.25}\smash{\footnotesize{\begin{tabular}[t]{l}[13], MSE = 0.235\end{tabular}}}}}%
    \put(0.692,0.50){\color[rgb]{0,0,0}\makebox(0,0)[lt]{\lineheight{1.25}\smash{\footnotesize{\begin{tabular}[t]{l}DP, MSE = 0.228\end{tabular}}}}}%
  \end{picture}%
\endgroup%

%% file: fig3_v4.pdf_tex
\begingroup%
  \makeatletter%
  \providecommand\color[2][]{%
    \errmessage{(Inkscape) Color is used for the text in Inkscape, but the package 'color.sty' is not loaded}%
    \renewcommand\color[2][]{}%
  }%
  \providecommand\transparent[1]{%
    \errmessage{(Inkscape) Transparency is used (non-zero) for the text in Inkscape, but the package 'transparent.sty' is not loaded}%
    \renewcommand\transparent[1]{}%
  }%
  \providecommand\rotatebox[2]{#2}%
  \newcommand*\fsize{\dimexpr\f@size pt\relax}%
  \newcommand*\lineheight[1]{\fontsize{\fsize}{#1\fsize}\selectfont}%
  \ifx\svgwidth\undefined%
    \setlength{\unitlength}{1082.34274292bp}%
    \ifx\svgscale\undefined%
      \relax%
    \else%
      \setlength{\unitlength}{\unitlength * \real{\svgscale}}%
    \fi%
  \else%
    \setlength{\unitlength}{\svgwidth}%
  \fi%
  \global\let\svgwidth\undefined%
  \global\let\svgscale\undefined%
  \makeatother%
  \begin{picture}(1,0.59404204)%
    \lineheight{1}%
    \setlength\tabcolsep{0pt}%
    \put(0,0){\includegraphics[width=\unitlength,page=1]{fig3_v4.pdf}}%
    \put(0.867,0.184){\color[rgb]{0,0,0}\makebox(0,0)[lt]{\lineheight{1.25}\smash{\small{\begin{tabular}[t]{l}ES\end{tabular}}}}}%
    \put(0,0){\includegraphics[width=\unitlength,page=2]{fig3_v4.pdf}}%
    \put(0.867,0.14){\color[rgb]{0,0,0}\makebox(0,0)[lt]{\lineheight{1.25}\smash{\small{\begin{tabular}[t]{l}DP\end{tabular}}}}}%
    \put(0.867,0.094){\color[rgb]{0,0,0}\makebox(0,0)[lt]{\lineheight{1.25}\smash{\small{\begin{tabular}[t]{l}ULA\end{tabular}}}}}%
    \put(0,0){\includegraphics[width=\unitlength,page=3]{fig3_v4.pdf}}%
    \put(0.362,-0.01){\color[rgb]{0,0,0}\makebox(0,0)[lt]{\lineheight{1.25}\smash{\begin{tabular}[t]{l}Beampattern MSE\end{tabular}}}}%
    \put(0.007,0.127){\color[rgb]{0,0,0}\rotatebox{90}{\makebox(0,0)[lt]{\lineheight{1.25}\smash{\begin{tabular}[t]{l}Achievable rate (bpcu)\end{tabular}}}}}%
    \put(0,0){\includegraphics[width=\unitlength,page=4]{fig3_v4.pdf}}%
    \put(0.29445724,0.32918398){\color[rgb]{0,0,0}\makebox(0,0)[lt]{\lineheight{1.25}\smash{\begin{tabular}[t]{l}Increasing $\mu$\end{tabular}}}}%
  \end{picture}%
\endgroup%

%% file: fig4_theNew_v2.pdf_tex
\begingroup%
  \makeatletter%
  \providecommand\color[2][]{%
    \errmessage{(Inkscape) Color is used for the text in Inkscape, but the package 'color.sty' is not loaded}%
    \renewcommand\color[2][]{}%
  }%
  \providecommand\transparent[1]{%
    \errmessage{(Inkscape) Transparency is used (non-zero) for the text in Inkscape, but the package 'transparent.sty' is not loaded}%
    \renewcommand\transparent[1]{}%
  }%
  \providecommand\rotatebox[2]{#2}%
  \newcommand*\fsize{\dimexpr\f@size pt\relax}%
  \newcommand*\lineheight[1]{\fontsize{\fsize}{#1\fsize}\selectfont}%
  \ifx\svgwidth\undefined%
    \setlength{\unitlength}{1071.20352173bp}%
    \ifx\svgscale\undefined%
      \relax%
    \else%
      \setlength{\unitlength}{\unitlength * \real{\svgscale}}%
    \fi%
  \else%
    \setlength{\unitlength}{\svgwidth}%
  \fi%
  \global\let\svgwidth\undefined%
  \global\let\svgscale\undefined%
  \makeatother%
  \begin{picture}(1,0.59071101)%
    \lineheight{1}%
    \setlength\tabcolsep{0pt}%
    \put(0,0){\includegraphics[width=\unitlength,page=1]{fig4_theNew_v2.pdf}}%
    \put(0.363,-0.009){\color[rgb]{0,0,0}\makebox(0,0)[lt]{\lineheight{1.25}\smash{\begin{tabular}[t]{l}Azimuth angle $\theta \, (^{\circ})$\end{tabular}}}}%
    \put(0.011,0.125){\color[rgb]{0,0,0}\rotatebox{90}{\makebox(0,0)[lt]{\lineheight{1.25}\smash{\begin{tabular}[t]{l}Beampattern $P(\theta)$ (dB)\end{tabular}}}}}%
    \put(0,0){\includegraphics[width=\unitlength,page=2]{fig4_theNew_v2.pdf}}%
    \put(0.7033,0.447){\color[rgb]{0,0,0}\makebox(0,0)[lt]{\lineheight{1.25}\smash{\footnotesize{\begin{tabular}[t]{l}ULA, MSE = 0.276\end{tabular}}}}}%
    \put(0,0){\includegraphics[width=\unitlength,page=3]{fig4_theNew_v2.pdf}}%
    \put(0.7033,0.553){\color[rgb]{0,0,0}\makebox(0,0)[lt]{\lineheight{1.25}\smash{\footnotesize{\begin{tabular}[t]{l}Desired\end{tabular}}}}}%
    \put(0.7033,0.518){\color[rgb]{0,0,0}\makebox(0,0)[lt]{\lineheight{1.25}\smash{\footnotesize{\begin{tabular}[t]{l}[13], MSE = 0.275\end{tabular}}}}}%
    \put(0.7033,0.482){\color[rgb]{0,0,0}\makebox(0,0)[lt]{\lineheight{1.25}\smash{\footnotesize{\begin{tabular}[t]{l}DP, MSE = 0.223\end{tabular}}}}}%
    \put(0,0){\includegraphics[width=\unitlength,page=4]{fig4_theNew_v2.pdf}}%
  \end{picture}%
\endgroup%